\documentclass[preprint,aps,singlecolumn,superscriptaddress,showpacs,floatfix,amssymb,amsmath]{revtex4-1}
\usepackage{epsfig}
\usepackage{graphics}
\usepackage{float} %for fixing pictures in certan place in text
\usepackage{array}
\usepackage{multirow}

\usepackage{yfonts}

\usepackage{graphicx}
\usepackage{amssymb}
\usepackage{mathrsfs}
\usepackage{bbm}
\usepackage{epsfig}
\usepackage{makecell}

\usepackage[usenames,dvipsnames]{color}

\usepackage{bbm}
\usepackage{color}

\newcommand{\be}{\begin{eqnarray}}
\newcommand{\ee}{\end{eqnarray}}

\begin{document}

\title{
Provenance  of classical Hamiltonian time crystals}
%\author{N.N.}
%\email{mail@address}
%\affiliation{University address}

% more complex case: 4 authors, 3 institutions, 2 footnotes
\author{Anton Alekseev}
\email{Anton.Alekseev@unige.ch}
\affiliation{Section of Mathematics, Universit\'e de G\`eneve \\ 2-4 rue du Li\`evre, Case postale 64
1211 Gen\`eve 4, Switzerland}
\author{Jin Dai}
\email{daijing491@gmail.com}
\affiliation{Nordita, Stockholm University,  Roslagstullsbacken 23, SE-106 91 Stockholm, Sweden}
\author{Antti J. Niemi}
\email{Antti.Niemi@su.se}
\affiliation{Institut Denis-Poisson CNRS/UMR 7013 - Universit\'e de Tours - Universit\'e d'Orl\'eans \\
Parc de Grammont, 37200 Tours, France}
\affiliation{Nordita, Stockholm University,   Roslagstullsbacken 23, SE-106 91 Stockholm, Sweden}
\affiliation{School of Physics, Beijing Institute of Technology, Haidian District, Beijing 100081, People's Republic of China}

\maketitle

{\bf
Classical Hamiltonian systems with conserved charges and those with constraints often describe dynamics on a pre-symplectic manifold. 
Here we show that a pre-symplectic manifold is also the proper stage to describe autonomous energy conserving Hamiltonian time crystals.  
We explain how the occurrence of a time crystal relates  to the  wider concept of spontaneously broken symmetries; 
in the case of a time crystal, the symmetry breaking takes place in a dynamical context.
We then analyze in detail two examples of time crystalline Hamiltonian dynamics.  The first example is a piecewise linear closed string, 
with dynamics determined by a Lie-Poisson bracket and Hamiltonian  that relates to membrane stability.  We explain how the Lie-Poisson
brackets  descents to a time crystalline  pre-symplectic bracket, and we show that the Hamiltonian dynamics supports two phases;
 in one phase  we have a time crystal and in  the other phase  
time crystals are absent.   The second example is a discrete Hamiltonian variant of  the Q-ball Lagrangian of time dependent 
non-topological solitons. We explain how a Q-ball becomes  a  time crystal, and we construct examples  of time crystalline Q-balls. }

\maketitle
\flushbottom

\section{Introduction}
\label{sec:intro}
 
 A classical Hamiltonian time crystal is an autonomous, time periodic solution of Hamilton's equation that is simultaneously 
a local minimum of the energy.  Accordingly a time crystal  spontaneously breaks the continuous 
time translation symmetry into discrete  time translations \cite{Wilczek-2012,Shapere-2012,Wilczek-2013}. (For  a review
see {\it e.g.} \cite{Sacha-2018,Nayak-2019}.) 
This symmetry breakdown is analogous to the way how an ordinary 
crystalline material breaks the group of continuous spatial translations into a discrete Bravais lattice.   

There are numerous  examples of periodically driven nonlinear oscillators, and many other kind of open and non-equilibrium  physical 
systems that display periodic time dependence. However,  these examples do not qualify as time crystals:
The breakdown of time translation symmetry is explicit and reflects verbatim the properties of the external forces and ambient 
interactions. Thus far veritable time crystalline dynamics \cite{Sacha-2018,Sacha-2016,khemani-2016,else-2016a,else-2016,else-2016b,Yao-2017,Nayak-2019,Elze-2019}
has been experimentally observed only in the context of certain externally 
driven non-equilibrium  spin chains \cite{zhang-2017,choi-2017}.
In the known examples a spin chain is subjected to an extrinsic, periodic driving force. This
prompts the spin chain to respond in a time periodic fashion, but  now the response comes with an intrinsic periodicity that is different 
from the period of the drive. 

Whenever an energy conserving, isolated Hamiltonian system has been found to 
display time periodic dynamics, such as  in the case of a harmonic oscillator or the Kepler problem,  the
motion can always be removed in a natural fashion, by a continuous deformation of the 
system towards its lowest available energy state.
%Whenever an energy conserving, isolated Hamiltonian system has been found to
%display time periodic dynamics, for example  in the case of a harmonic oscillator and in the Kepler problem,  the 
%periodic time dependent motion is always expunged when the system reaches its lowest available energy state. 
Indeed, it is widely 
thought that in the case of a closed autonomous Hamiltonian system, any kind of time crystalline 
dynamics is excluded \cite{bruno-2013,watabane-2014}. 
This conclusion is grounded on the structure of  the textbook Hamilton's equation
\begin{equation}
\begin{split}
& \frac{dq^a}{dt} = \{ q^a , H \} = \frac{\partial H}{\partial p^a}
\\
& \frac{dp^a}{dt} = \{ p^a , H \} =  - \frac{\partial H}{\partial q^a}
\end{split}
\label{pqH}
\end{equation}
On a compact, closed manifold a minimum energy configuration is also a critical point of the Hamiltonian $H$. Thus, at the energy minimum
the right hand sides of (\ref{pqH})
vanish,  which implies that the left hand sides must vanish as well. As a consequence a trajectory that is a minimum of
$H$  must be  time independent and one concludes that Hamilton's equation (\ref{pqH}) can not support any time crystalline solution. 

However, we have found that there is a systematic way to evade these  Hamiltonian  {\it No-Go} arguments. This is exemplified by the following
simple scenario: 
Suppose that $h = h(x,y)$ is a smooth real valued function defined on a compact closed manifold. For example
we may take the manifold to be  
the Riemann sphere {\it i.e.} the plane $\mathbb R^2$ plus a point at infinity.  Assume that the function has only 
non-degenerate maxima and minima on the sphere, these are located at  critical points 
$\partial_x h = \partial_y h =0$.  % for  example if $h = h(x,y)$ is the height function the maxima and minima are at the north and south poles.
Now consider another  smooth real valued function $g(x,y)$ and
introduce a condition such as
\begin{equation}
g(x,y) = c
\label{gxy}
\end{equation} 
Whenever  $c$ is a regular value of $g$ the solution defines a curve on the sphere. Generically, for a given value
of $c$ we  do not expect that the  curve passes through any of the  critical points of $h(x,y)$.  Thus, for a generic given $c$ the
minimal values that $h(x,y)$ attains along the curve $g(x,y) = c$ are not among its critical points on the sphere. It 
is even possible that this is the case for all those values of $c$ that are allowed by the structure of   $g(x,y)$ 

Recently, explicit examples of  classical Hamiltonian time crystals have been presented \cite{Dai-2019}. The examples
go around the {\it No-Go} arguments in the manner that we have outlined: %notwithstanding that  their dynamics is governed by a Lie-Poisson bracket:
They are Hamiltonian systems with conserved quantities, but  the  numerical values of the conserved charges
are constrained.  This causes the Hamiltonian system to  become time crystalline, 
with a spontaneously broken time translation symmetry.  In particular,  the ensuing phase spaces are not
symplectic manifolds. Instead, the time crystalline dynamics takes 
place on a more general pre-symplectic manifold \cite{Marsden-1999}  where  the {\it No-Go} arguments are circumvented.

We  start and trace the provenance of Hamiltonian time crystalline dynamics 
to the geometry of the phase space. %We  argue that the phase space needs to be a pre-symplectic manifold.  
We explain in a  general context why pre-symplectic geometry is necessary for the emergence of time crystalline dynamics, 
in the case of Hamilton's equation with conserved charges.  We then  construct in detail two examples of Hamiltonian time crystals,  
both examples  have their origin and motivation in familiar field theoretical models.

\section{Hamiltonian time crystals}
\label{sec:intro}

%\subsection{Elemental considerations} 

Hamilton's equation describes energy conserving dynamics on a $2N$ dimensional 
symplectic manifold $\mathcal M$;  for background on geometric mechanics see {\it e.g.}  \cite{Marsden-1999}.
The manifold is equipped with a closed and non-degenerate two-form 
\begin{equation}
\begin{split}
& \Omega = \Omega_{ab} d\phi^a \wedge d\phi^b \\ 
d & \Omega = 0
\end{split}
\label{Omega}
\end{equation}
where $\phi^a$ ($a=1,...,2N$) are generic local coordinates on $\mathcal M$. For pedagogy   
we prefer to use local coordinate representation even though all our relations are 
coordinate independent. Hamilton's equation is
\begin{equation}
\Omega_{ab} \frac{ d\phi^b}{dt} = \partial_a H
\label{phiH0}
\end{equation}
where the Hamiltonian $H(\phi)$ models  the mechanical  free energy, it is assumed to be  a smooth real valued function 
on $\mathcal M$.  
The solutions $\phi^a(t)$ are non-intersecting trajectories on $\mathcal M$, they are uniquely specified by the initial values $\phi^a(0)$.
 The  inverse  of the matrix $\Omega_{ab}$ defines  the Poisson brackets
\begin{equation}
\{ \phi^a , \phi^b \} = \Omega^{ab}(\phi)
\label{PBphi}
\end{equation}
so that  (\ref{phiH0})  becomes
\begin{equation}
\frac{ d\phi^a}{dt} \ = \ \{ \phi^a , H \} \ = \ \Omega^{ab} \partial_b H
\label{phiH}
\end{equation}
Darboux theorem ensures that there is a  local coordinate transformation on the manifold  $\mathcal M$   
such that the $\phi^a$ become the ($p^a,q^a$) that are equipped with their 
canonical Poisson brackets,  and Hamilton's equation  (\ref{phiH}) acquires  the familiar form  (\ref{pqH}).

\vskip 0.2cm

We define a time crystal  to be a minimum energy 
solution of Hamilton's equation (\ref{phiH0}), (\ref{phiH}) with a non-trivial $t$-dependence that we assume is periodic
$\phi^a(t+T) = \phi^a(t)$.%\footnotemark\footnotetext{Quasiperiodic minimum energy trajectories can also exist. They also break 
%time translation symmetry spontaneously and their properties can be analyzed similarly. For clarity, here we consider only genuine time periodic 
%time crystals.} 

\vskip 0.2cm
We search for time crystals in Hamiltonian systems with symmetries.  
Noether's theorem states that a symmetry gives rise  to a
conservation law and we denote the pertinent conserved charges $G_i(\phi)$ ($i=1,...,n \leq N$).
Their Poisson brackets with the Hamiltonian $H(\phi)$ vanish,
 \[
 \{ H, G_i\} = \frac{d G_i}{dt} \  = \ 0
 \] 
Furthermore, the Poisson brackets of the $G_i$ closes with a  Lie algebra structure
\begin{equation}
\{ G_i, G_j\} = {f_{ij}}^k G_k
\label{LieG}
\end{equation}
of the symmetry group. 

We assign numerical values  $G_i(\phi(0)) = g_i$ to the conserved charges.
The $ g_i$ are  regular values of the $G_i(\phi)$,  they are determined in terms of the initial conditions $\phi^a(0)$ of Hamilton's equation. 
The level sets  $G_i(\phi(0)) = g_i$ foliate the symplectic manifold $\mathcal M$, and we specify them
by conditions
\begin{equation}
\mathcal G_i^g (\phi) \ = \ G_i(\phi) - g_i = 0
\label{Gg}
\end{equation}
For each $g_i$  the condition (\ref{Gg}) specifies a  submanifold of $\mathcal M$ 
that we denote by $\mathcal M_g$.

The Poisson brackets of (\ref{Gg})  are 
\begin{equation}
\{ \mathcal G^g_i , \mathcal G^g_j \} = {f_{ij}}^k \, \mathcal G^g_k + {f_{ij}}^k g_k
\label{G+g}
\end{equation}
where the matrix 
\begin{equation}
\gamma_{ij} (g) = {f_{ij}}^k g_k
\label{gamma}
\end{equation}
has a rank $s\leq n$ that in general depends on the values $g_i$. Following  Dirac \cite{Dirac} we regard 
(\ref{Gg}) as a combination of $(n-s)$ first class conditions and $s$ second class conditions. 
The first class conditions correspond to those combinations of
$G_i(\phi)$ that define  the kernel of $\gamma_{ij} (g)$ and the second class conditions span the image of $\gamma_{ij}(g)$.

For  each regular values $ g_i$ in (\ref{Gg}) 
we restrict  the non-degenerate symplectic two-form (\ref{Omega})  to the corresponding  submanifold $\mathcal M_g$
\begin{equation}
\Omega_{|_{\mathcal M_g}} \ \equiv \  \omega^g = \omega^g(\phi)_{ab} d\phi^a \wedge d\phi^b
\label{ome}
\end{equation}
The two-form $\omega^g$  is closed but in general the matrix
$ \omega^g_{ab}$ is degenerate with a  ($n-s$) dimensional  kernel.
% that coincides with the submanifold spanned by the ($n-s$)  first 
% class conditions of (\ref{Gg}).  The $s$-dimensional image of $ \omega^g_{ab}$ 
% coincides with the submanifold that is spanned by the $s$ second class conditions of  (\ref{Gg}).   
Accordingly,  whenever $n-s \not= 0$ the submanifold $\mathcal M_g$ that we equip with the closed two-form (\ref{ome}), 
is  not a symplectic manifold  but a pre-symplectic manifold. That is, simply a manifold with a closed two-form.  

In the following we assume that  the  physical circumstances are such that for all regular values $g_i$ of interest in  (\ref{Gg}) we have
$n-s \not= 0$ so that  the  corresponding manifolds $\mathcal M_g$ are pre-symplectic. 
Since the {\em No-Go} arguments \cite{bruno-2013,watabane-2014} assume that Hamiltonian dynamics takes place on a symplectic manifold,  those
arguments no longer apply. Thus, in the case of a pre-symplectic manifold  $\mathcal M_g$  the existence of a time crystalline solution 
to Hamilton's equation can not be excluded.

We note that there are many Hamiltonian dynamical systems with conserved charges. However,  not all of them can support a 
time crystal. The existence of a time crystal depends on the way how the regular values $g_i $ in (\ref{Gg}) are  distributed into 
subsets $\{g_i\} \subset \mathbb R^n$.  In general there  can be  multiple disconnected subsets,  
and  each connected  component  pertains to a specific physical scenario.  
A necessary condition for a given subset $\{g_i\}$ to support time crystalline  dynamics is,  that this subset can not be  
path connected to a stationary point of the Hamiltonian with a lower energy,  
in a manner that is dictated by the specifics of the physical scenario.\footnotemark\footnotetext{
This excludes examples such as the Kepler problem where the angular momentum can continuously decrease until it vanishes,
and the harmonic oscillator where we can continuously diminishing the energy. } The concrete examples that we present in the sequel,
elucidate this point.

To reveal the actual presence of a time crystal
we need to locate the minimum value of the Hamiltonian $H(\phi)$.  Since the $G_i(\phi)$ are conserved it suffices to restrict the search of the 
minimum to the submanifolds $\mathcal M_g$ of interest. 
For this we account for (\ref{Gg}) in terms of  Lagrange multipliers
$\lambda^i$ and extend the Hamiltonian $H(\phi)$ into
\begin{equation}
H \to H_\lambda = H + \lambda^i (G_i - g_i)
\label{crit} 
\end{equation}
The  Lagrange multiplier theorem  \cite{Marsden-1999} states that on a given  submanifold (\ref{Gg})  the minimum  value $\phi^a_{cr}$ 
of the Hamiltonian $H(\phi)$ coincides  with  a critical point ($\phi^a_{cr}, \lambda_{cr}^i$) of the Hamiltonian function $H_\lambda(\phi)$. 
Thus  the minimum value of $H(\phi)$ on $\mathcal M_g$ is obtained as a solution of
\begin{equation}
\left\{ 
\begin{split}
%\label{lambda}
\frac{\partial H}{\partial \phi^a}_{| \phi_{cr} } &  =   - \lambda_{cr}^i \frac{ \partial G_i }{\partial \phi^a}_{| \phi_{cr} } \\ 
G_i(\phi_{cr})   & =  g_i 
\end{split} 
\right.
\label{Glambda}
\end{equation} 
Accordingly, we search for a time crystal using the following steps:

\vskip 0.3cm

\noindent
$\bullet ~$  From the equations (\ref{Glambda})  we first solve for  the minima $\phi^a_{cr}$ of  $H(\phi)$
on the submanifolds $\mathcal M_g$  of interest. Here the set of $\mathcal M_g$  includes all the level surfaces of the conserved charges that 
correspond to  the physical scenario.

\vskip 0.1cm
\noindent
$\bullet ~$ We then continue and solve (\ref{Glambda})  for the corresponding values  
$\lambda^i_{cr}$  in terms of $\phi^a_{cr}$. 

\vskip 0.3cm

\noindent 
%$\bullet ~$  
Whenever $\lambda^i_{cr}(\phi_{cr})  \not=0$ 
the  minimum energy solution $\phi^a_{cr}$ can be employed as an initial value to a time crystalline solution of 
Hamilton's equation (\ref{phiH}).
In the case of a time crystal,  Hamilton's equation then becomes 
\begin{equation}
\begin{split}
& \frac{d\phi^a}{dt} \  =    \
 %\Omega^{ab}\partial_b H  =  
 - \Omega^{ab} \lambda^i_{cr}  \frac{\partial G_i}{\partial \phi^b}  \not=0  \\ 
& \phi^a(0)  \ = \   \phi^a_{cr}
 \end{split}
\label{creq}
\end{equation}

Note that the  Lagrange multipliers $\lambda^i_{cr}$  are $t$-independent,  their values for all $t$ 
are determined by  (\ref{Glambda}) in terms of the initial values $\phi^a_{cr}$. This follows immediately, since    
both $H(\phi)$ and $G_i(\phi)$ are by construction $t$-independent along {\it any} Hamiltonian trajectory.

The emergence of a time crystal is a manifestation of the general phenomenon of spontaneous symmetry breaking: 
A time crystal  describes a time dependent minimum 
energy symmetry transformation of the Hamiltonian $H(\phi)$ that is  generated by the linear combination 
\[
G_\lambda(\phi) \equiv \lambda^i_{cr} G_i(\phi)
\]
Thus a time crystal breaks the full symmetry group  of conserved charges 
(\ref{LieG}) into an abelian U(1) symmetry transformation.  
We remark that in general the ensuing motion (\ref{creq}) is  quasi-periodic, but here we assume it to be periodic $\phi^a(t+T) = \phi^a(t)$
and the period $T$ is specified by $\lambda^i_{cr}(\phi_{cr})$.

\vskip 0.2cm
It is apparent that the present remarks are merely an invitation for a judicious mathematical investigation, and we  propose that the methods of 
equivariant Morse theory \cite{Wasserman-1969,Austin-1995,Nicolaescu-2011} can be adopted to develop a mathematical framework for
understanding Hamiltonian time crystals;  we plan to return to this in a future research and we now proceed to exemplify our general 
remarks by a detailed analysis of  two examples where time crystalline dynamics appears in a familiar physical context.

%\section{Examples}

\section{Example 1: Time crystals and closed discrete strings}

%\subsubsection{Reduced Yang-Mills-Chern-Simons Hamiltonian}

In our first example we  follow \cite{Dai-2019}  and analyze  time crystalline dynamics  in  the context of a Hamiltonian system with 
time evolution determined by a Lie-Poisson bracket \cite{Marsden-1999}.  The  Hamiltonian function we use has been 
originally introduced in \cite{Alexeev-2000,Monnier-2005,Alekseev}, in
connection of membrane stability analysis. 
%In the  Appendix we also motivate the Hamiltonian function, 
%using a reduction of axion coupled Yang-Mills theory.

We start by explaining  how the Lie-Poisson structure  of \cite{Dai-2019}  fits in our general framework:
A Lie-Poisson bracket commonly describes the way how a Poisson manifold, {\it i.e.} a manifold 
that is equipped with a Poisson bracket, becomes foliated 
into  symplectic leaves. Each leaf is a symplectic manifold,  it supports Hamiltonian dynamics that is 
restrained on the given leaf.  

We consider a four dimensional phase space with Darboux coordinates ($q_1,q_2,p_1,p_2$)
that we combine into complex coordinates 
\begin{equation}
\begin{split}
z_1 = \frac{1}{2} (p_i + i q_1)/2  \\
z_2 = \frac{1}{2} (p_2 +i q_2)/2 
\label{z12}
\end{split}
\end{equation}
with  Poisson brackets
\begin{equation}
%\begin{split}
\{ z_i , z_j^\star \}  = i \delta_{ij} \ \ \ \ \& \ \ \ \ \  \{ z_i , z_j \}  = \{ z_i^\star , z_j^\star \} \ = \ 0
%\end{split}
\label{z-bra}
\end{equation}
We introduce the Pauli  matrices 
\[
\sigma^1 = \left( \begin{matrix} 0 & 1 \\ 1 & 0 \end{matrix} \right)  \ \ \ \ \  \sigma^2 = \left( \begin{matrix} 0 & -i  \\ i & 0  \end{matrix} \right)   
\ \ \ \ \  \sigma^3 = \left( \begin{matrix} 1 & 0  \\ 0 & -1 \end{matrix} \right)  
\]
to define a vector with three real components
\begin{equation} 
n^a = -\frac{1}{2} \left( z_1^\star \,  , \,  z_2^\star \right) \sigma^a \left( \begin{matrix} z_1 \\ z_2 \end{matrix} \right) 
\label{n}
\end{equation}
and so that
\begin{equation}
\mathbf n \cdot \mathbf n = \frac{1}{4} ( |z_1|^2 + |z_2|^2)^2 \ = \ r^4
\label{nl}
\end{equation}
The $n^a$ obey the SU(2) Lie-Poisson brackets
\begin{equation}
\{ n^a , n^b \} = \epsilon^{abc} n^c
\label{n-bra}
\end{equation}
and the length (\ref{nl}) is preserved by the action of the $n^a$
\begin{equation}
\{  n^a , \mathbf n \cdot \mathbf n \} = 0
\label{n-r}
\end{equation}
The phase space  (\ref{z12}) is a model space of SU(2) representations, and different values of $r$ correspond to different representations.
 We identify the symplectic structures of the SU(2) coadjoint orbits  in terms of local coordinates
\begin{equation}
\left( \begin{matrix} z_1 \\ z_2 \end{matrix} \right) \ = \ r \left( \begin{matrix} \cos\frac{\theta}{2} \, e^{ i (\chi + \phi)/2 }  \\ 
\sin\frac{\theta}{2} \, e^{ i (\chi - \phi)/2 }  \end{matrix} \right) 
\label{zomega}
\end{equation}
This yields
\begin{equation}
\mathbf n = \left( \begin{matrix} n_1 \\ n_2 \\ n_3 \end{matrix} \right) \ = \ r^2 \left( \begin{matrix} \cos\phi \sin\theta  \\ \sin\phi \sin\theta \\ \cos\theta \end{matrix} \right) 
\label{nproj}
\end{equation}
and for the symplectic two-form of (\ref{z-bra})  we get
\[
\Omega = i d z_1^\star \wedge d z_1 + id z_2^\star \wedge d z_2
\ = \  r d\chi \wedge dr +  r \cos\theta \, d \phi \wedge dr + \frac{r^2}{2} d\phi \wedge d \! \cos\theta 
\]
from which we read the following Poisson brackets
\begin{eqnarray}
%\begin{split}
\label{s-pb1}
 \{ r , \chi \}  & = & \frac{1}{r}  \\
\label{s-pb2}
\{ \cos \theta , \chi \}  & =  & - \frac{2}{r^2} \cos\theta \\
\label{s-pb3}
 \{ \cos\theta , \phi \}  & = & \frac{2}{r^2} \\
\label{s-pb4}
\{ r, \cos\theta \} =  \{ r, \phi \} & = & \{ \chi ,  \phi \} = 0
%\end{split}
\end{eqnarray} 
The coordinates in the {\it r.h.s.} of (\ref{zomega})  are simply spherical coordinates on $\mathbb R^4$, the  ($\theta, \phi, \chi$)  are angular coordinates that
describe  the spheres $\mathbb S^3$ that foliate $\mathbb R^4$ with radii $r^2$. The Hopf map $\mathbb S^3  \to \mathbb S^2$ identifies  ($\theta, \phi$) as the 
latitude and longitude angles of a two-sphere, 
and  $\chi$ is the coordinate of the remaining $\mathbb S^1$. The two-spheres are the orbits of 
SU(2) representations,  and each two-sphere is equipped with a symplectic two-form that corresponds to the Poisson bracket (\ref{s-pb3}), with $\cos\theta$ and
$\phi$ a canonical pair.

We proceed to describe a physical scenario where  $N$ such vectors $\mathbf n_i$ appear as dynamical degrees of freedom, 
each equipped with its own Lie-Poisson bracket (\ref{n-bra}) \cite{Dai-2019}.  
For this we interpret the vectors as links
that connect the $N+1$ vertices $\mathbf x_i $ of a  piecewise linear polygonal string  in $\mathbb R^3$
\begin{equation}
\mathbf n_i = \mathbf x_{i+1} - \mathbf x_i 
\label{nx}
\end{equation}
Since
\[
\{ \mathbf n_i , \mathbf n_k \cdot \mathbf n_k \} \ = \ 0 \ \ \ \ \ \ {\rm for ~ all} \ i,k
\]
the lengths of the links remain intact during time evolution, 
whenever the Hamiltonian function depends only on the vectors $\mathbf n_i$ which we assume to be the case. 
For convenience we set all  the link lengths to have the equal value $|\mathbf n_i | = 1$.

With  $H(\mathbf n)$ a Hamiltonian function,  the Lie-Poisson bracket (\ref{n-bra}) yields the following Hamilton's equation
\begin{equation}
\frac{\partial \mathbf n_i}{\partial t} =  \{ \mathbf n_i , H(\mathbf n) \} = - \mathbf n_i  \times \frac{\partial H}{\partial \mathbf n_i}
\label{YMeq}
\end{equation}
and we proceed to reveal its time crystalline dynamics. 

To introduce the 
conserved charges (\ref{Gg}) we consider the vector 
\begin{equation}
\mathbf G \ = \ \sum\limits_{i=1}^N \mathbf n_i % \ = \ \mathbf x_{N+1} - \mathbf x_1 % = 0 \ \  \ \ \ \ {\rm with} \ \  \mathbf x_{N+i} = \mathbf x_i
\label{G}
\end{equation}
Its components obey the Poisson brackets
\begin{equation}
\{ G^a , G^b \} = \epsilon^{abc} G^c 
\label{gauss-1}
\end{equation}
and  we choose a Hamiltonian such that
\begin{equation}
\{ H (\mathbf n) , \mathbf G \} = 0
\label{gauss-2}
\end{equation}
The $\mathbf G$  are the conserved charges of interest.  The following is then 
an example of the condition (\ref{Gg}): We define the manifolds $\mathcal M_g$  by
\[
G^a = g^a \ \equiv \ (x^a_{N+1} - x^a_1) 
\]
and the matrix $\gamma_{ij}$ in (\ref{gamma}) is
\[
\gamma_{ij} \ \sim \ \epsilon^{abc} (x^c_{N+1} - x^c_1)
\]

We now specify the physical scenario that is of interest to us: We describe the dynamics of a  closed string, 
and for this we set $\mathbf x_{N+1} = \mathbf x_1$.
Thus (\ref{Gg}) becomes
\begin{equation}
\mathbf G \ = \ \sum\limits_{i=1}^N \mathbf n_i   = 0
\label{G0} 
\end{equation}
The entire algebra (\ref{G+g}) of the  conserved charges is first class.

In line with our general formalism we  
introduce the Lagrange multiplier $\boldsymbol{\lambda}$ so that   the Hamiltonian (\ref{crit}) in the present case is 
\[
H_{\boldsymbol \lambda} = H(\mathbf n) + {\boldsymbol \lambda}\cdot \mathbf G
\]
For a time crystal, the equation (\ref{creq}) yields us the following
\begin{equation}
\frac{\partial \mathbf n_i}{\partial t} \ = \  - {\boldsymbol \lambda}_{cr} \times \mathbf n_i  
\label{l-crit}
\end{equation}
where the Lagrange multiplier ${\boldsymbol \lambda}_{cr}$ is evaluated at the minimum value of the Hamiltonian, 
\begin{equation} 
{\boldsymbol \lambda}_{cr}  =  - \frac{ \partial H}{\partial \mathbf n_i}_{| \mathbf n_{min}}
\label{lambda-eq}
\end{equation}
If a solution with ${\boldsymbol \lambda}_{cr}\not= 0 $ exist we have a time crystalline closed string that 
rotates as a rigid body.  The rotation axis points in
the direction of  ${\boldsymbol \lambda}_{cr}$ and the magnitude of the angular velocity is given by the length 
$|{\boldsymbol \lambda}_{cr}|$.
Note that the {\it r.h.s.} involves the index $i=1,...,N$ that is absent in the {\it l.h.s.} Thus (\ref{lambda-eq}) is  a very stiff condition on the
shape of the time crystalline  closed string.

\vskip 0.2cm

We proceed to analyze in detail two such time crystalline closed strings, with $N=3$ and $N=4$ vertices. 
For the Hamiltonian function, we follow \cite{Alexeev-2000,Monnier-2005,Alekseev} and  select   
\begin{equation}
H = H_1 + c H_2 = \sum\limits_{i=1}^N | \mathbf n_i \times \mathbf n_{i+1} |^2 + c 
\sum\limits_{i=1}^{N}  \mathbf n_i \cdot (\mathbf n_{i+1} \times \mathbf n_{i+2}) \ \ \ {\rm with} \ \ \ \mathbf n_{N+i} = \mathbf n_i
\label{Hym}
\end{equation}

\subsection{Three-vertex model}

For $N=3$ the closed string constraint (\ref{G})  states that the variables $\mathbf x_1, \mathbf x_2, \mathbf x_3$ are the vertices of an equilateral triangle in 
$\mathbb R^3$. We can take  $\mathbf x_1, \mathbf x_2, \mathbf x_3$ to lie on the $xy$-plane,  with sites $|\mathbf x_{i+1} - \mathbf x_i| = 1$.  
With the initial choice 
\[
\begin{split}
\mathbf n_1 & = (1,0,0)  \\
\mathbf n_2 & =   ( - \frac{1}{2} ,  \frac{\sqrt{3}}{2} , 0 ) \\
\mathbf n_3 & = ( - \frac{1}{2} , - \frac{\sqrt{3}}{2} , 0 )
\end{split}
\]
a direct substitution of  (\ref{Hym}) into (\ref{YMeq}) gives %for (\ref{creq})
\[
\frac{\partial \mathbf n_i}{\partial t} =   -  {\boldsymbol \lambda}_{cr} \times \mathbf n_i
\]
where ${\boldsymbol \lambda}_{cr}$  coincides with the symmetry axis of the triangle,
\[
{\boldsymbol \lambda}_{cr} \ = \ \frac{\sqrt{3}}{2}  ( 0 , 0 , c)  
\] 
Thus,  whenever $c\not=0$ we have a time crystal that describes an equilateral triangle that rotates around its symmetry axis with an angular velocity that 
is linearly proportional to the parameter $c$ in (\ref{Hym}). Note that for $c=0$ only the first term in (\ref{Hym}) is present. For this $c$-value
${\boldsymbol \lambda}$ vanishes; there is no time crystal if only $H_1(\mathbf n)$ is present.

The present example is  a very simple realization of the general result (\ref{creq}), as the
equilateral triangle can not change  its shape and the Hamiltonian (\ref{Hym}) has only a single 
value which is simultaneously the minimum and maximum 
of the available energy.

\subsection{Four-vertex model} 

The Hamiltonian  (\ref{Hym})  can be readily extended to a closed string with 
more than $N=3$ variables. As an example we consider  a closed polygonal string with $N=4$ vertices, 
and for convenience we 
take $|\mathbf x_{i+1}- \mathbf x_i|=1$. Geometrically, we may view
the $\mathbf x_i$ as the vertices of a tetrahedron in $\mathbb R^3$ with four equal length edges.  
Up to an additive  constant 
the Hamiltonian (\ref{Hym}) is
\begin{equation}
H = H_1 + H_2 = - \sum\limits_{i=1}^4 (\mathbf n_i \cdot \mathbf n_{i+1})^2 + c \, \mathbf n_1 \cdot (\mathbf n_2 \times \mathbf n_3)  \ \ \ \ \ ({\rm with} ~ \mathbf n_5 = \mathbf n_1)
\label{N4H}
\end{equation}
To construct a  time crystal, we first minimize  the energy (\ref{N4H}) on the constraint manifold (\ref{G}) for different parameter values $c$. 
%
%
%                   FIGURE_1--3
%
%
%
\begin{figure}[h]
\centering % \begin{center}/\end{center} takes some additional vertical space
%\includegraphics[width=.45\textwidth,trim=0 380 0 200,clip]{figure-1.pdf}
%\hfill
\includegraphics[width=.50\textwidth,origin=c,angle=0]{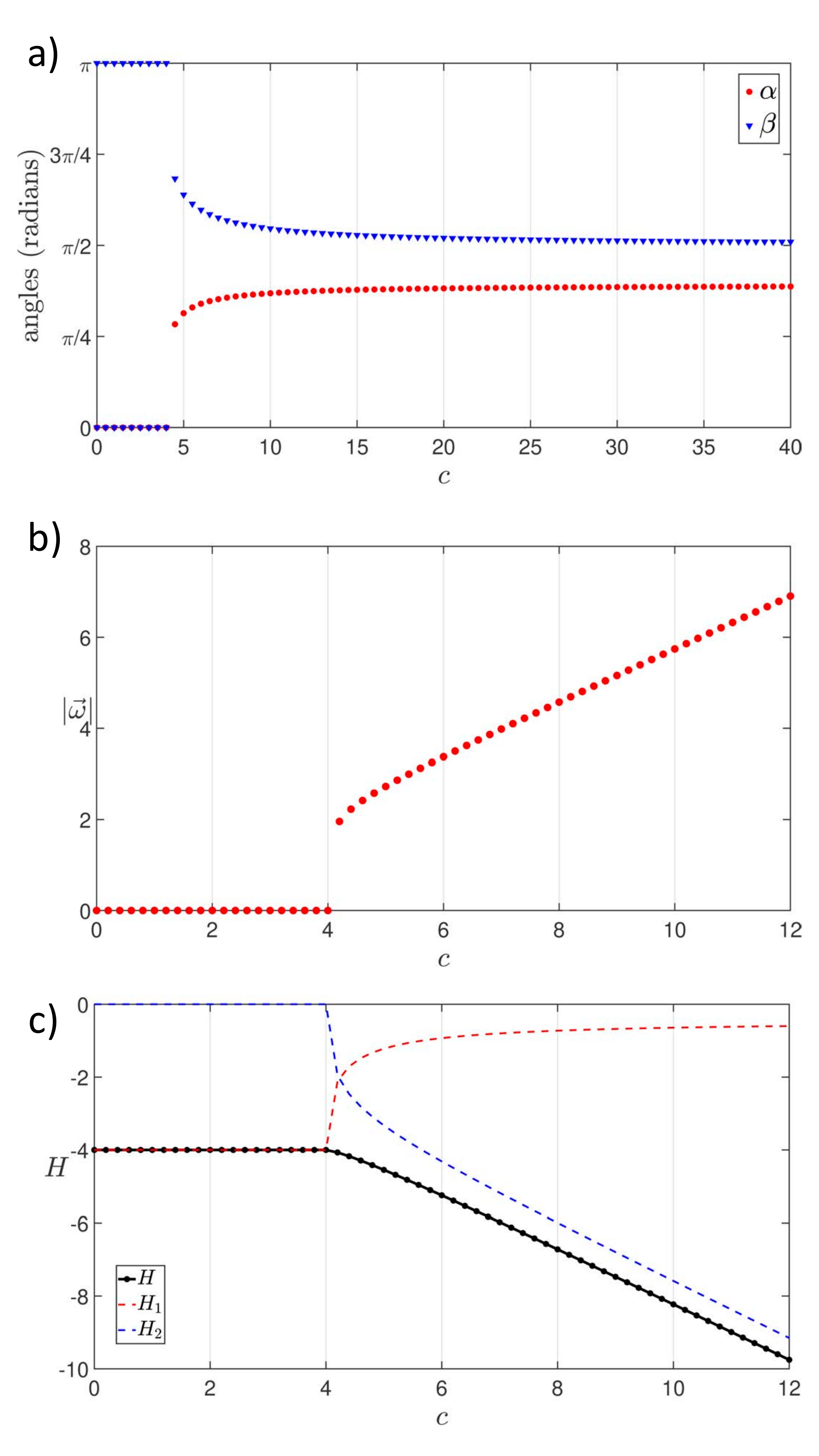}
% "\includegraphics" is very powerful; the graphicx package is already loaded
\caption{\label{fig-1} Panel a) The minimum energy values of the angles (\ref{ab}) as a function of the parameter $c$ in (\ref{N4H}).  
The asymptotic large-$c$ values ($\alpha,\beta$) = ($\arccos(1/3), \pi/2$) correspond to a tetragonal disphenoid.Panel b)
The angular velocity of the time crystal with Hamiltonian (\ref{N4H})  as a function of the parameter $c$.
Panel c)  The total energy $H$ and the individual contributions $H_1$ and $H_2$ in  (\ref{N4H}), as a function of parameter $c$ for the time crystal.}
\end{figure}
Since the four edges have equal length, $H$ is a function of the following two conformational angles,
\begin{equation}
\begin{split}
& \alpha  = \  \pi - \arccos ( \mathbf n_1 \cdot \mathbf n_4) \hspace{2.9cm}  \ \in \ (0,\pi]  \\
%\label{a}
%\end{equation}
%\begin{equation}
& \beta \ =  \ \pi - \arccos\left( \frac{ ( \mathbf n_1 -  \mathbf n_4) \cdot ( \mathbf n_3 - \mathbf n_2) }
{  |\mathbf n_1 -  \mathbf n_4 |  |\mathbf n_3 -  \mathbf n_2| } \right) \ \in \ (0,\pi]
\end{split}
 \label{ab}
 \end{equation} 
Here $\alpha$ is the bond angle $\angle ( \mathbf x_4  \mathbf x_1  \mathbf x_2)$ and $\beta$ is the dihedral angle 
between the two planes $ <\!\!\mathbf x_1 \mathbf x_2 \mathbf x_4\!\!>$  and $<\!\! \mathbf x_2 \mathbf x_3 \mathbf x_4\!\!>$. 
Accordingly, we proceed and express the Hamiltonian (\ref{N4H}) in terms of (\ref{ab}), and then search for the minimum of
$H(\alpha,\beta)$ with different  parameter values $c$.  For  energy minimization we  use a two dimensional adaptive grid algorithm.

\vskip 0.3cm
\noindent
We find that there is a critical value $c = 4.0 $ with the following properties:

\vskip 0.1cm

 When $c<4.0$  the minimum of $H(\alpha,\beta)$ is doubly degenerate, both ($\alpha,\beta$) = ($0,0$) and  ($\alpha,\beta$) = ($0,\pi$) are minima of the energy 
and at these minima we have $H=H_1=H_2=0$.   Geometrically, in both energy minima the  four link vectors $\mathbf n_i$ are all aligned with each other
in $\mathbb R^3$.  In the case of ($\alpha,\beta$) = ($0,0$) 
the neighboring vectors $\mathbf n_1$ and $\mathbf n_{2}$ are mutually parallel and opposite to the mutually parallel $\mathbf n_{3}$ and $\mathbf n_{4}$.
In the case of  ($\alpha,\beta$) = ($0,\pi$)  the vectors
$\mathbf n_1$ and $\mathbf n_{3}$ are mutually parallel, and opposite to the direction of the mutually parallel $\mathbf n_{2}$ and $\mathbf n_{4}$.
When we substitute in (\ref{YMeq}) we find no motion, neither of the two energy minima corresponds to a time crystal and we conclude that
for $c<4.0$ there is no time crystal.

When $c=4.0$ we observe an abrupt change in the  values of ($\alpha,\beta$).
Figure \ref{fig-1} Panel a) shows how  the values of ($\alpha,\beta$) jump from the
$c < 4.0$ minimum energy values ($\alpha,\beta\, $)=($\, 0,0$), ($0,\pi$) to  ($\alpha,\beta$)$\, \approx\, $($0.87, 2.18$) radians for $c = 4.0+$,
and Panel b) shows how the angular velocity  also changes abruptly. 
We observe a change in the shape of the minimum energy polygonal string, for $c < 4.0$ the 
distances $d_{13} = |\mathbf x_3 - \mathbf x_1| $ and $d_{24} =  |\mathbf x_4 - \mathbf x_2|$ have the values  $d_{13}=0$ and $d_{24} = 2$, but for
$c =4.0+$ these values are $d_{13}=d_{24} = 0.8$.

When $c>4.0$ we find that the minimum energy configuration is a time crystal, it rotates with an angular 
velocity $|{\boldsymbol \omega}|$ that increases when $c$ increases; see Figure \ref{fig-1} Panel b).  
At the same time, the total  energy decreases as a function of $c$, 
as shown in Figure  \ref{fig-1} Panel c).
Asymptotically, for large values of $c$, both the 
minimum value of the energy and  the angular velocity
depend on $c$ linearly.  For the conformational angles the large-$c$ limiting values are 
($\alpha,\beta$) = ($\arccos(1/3), \pi/2$) and thus, in the  large-$c$ limit the four vertices $\mathbf x_1 , ... , \mathbf x_4$ approach the vertices
of the space filling tetragonal disphenoid with $d_{13}=d_{24}=2/\sqrt{3}$; see Figure \ref{fig-1} Panel a).

 \section{Example 2: Time crystalline Q-balls}

%\subsubsection{Reduced Q-ball Hamiltonian}

\vskip 0.2cm
Our second example is based on the Q-ball model of \cite{Coleman-1985}; for surveys see  \cite{Lee-1992,Volkov-2008}. 
There is one complex field $\varphi(\mathbf x,t)$ and the 
Hamiltonian form of the (relativistic) action is
\begin{equation}
% \int dt d^D\! x \,  \left\{ \pi \dot \varphi + \pi^\star {\dot \varphi}^\star  - H (\pi, \varphi) \right \} =
  \int dt d^D\! x  \left\{ \pi \dot \varphi + \pi^\star {\dot \varphi}^\star - \pi \pi^\star - |\nabla \varphi |^2  - U(|\varphi |) \right\} 
  =   \int dt d^D\! x  \left\{ \pi \dot \varphi + \pi^\star {\dot \varphi}^\star - H(\pi,\varphi) \right\}
%- \frac{1}{2} m^2 |\phi |^2 + \frac{1}{4} \lambda  |\phi |^4 - \frac{1}{8} \sigma |\phi|^6 \right\}
\label{SQ1}
\end{equation}
The non-vanishing Poisson brackets are
\[
\{ \varphi (\mathbf x), \pi(\mathbf y)  \} = \{ \varphi^\star (\mathbf x), \pi^\star (\mathbf y) \} = \delta(\mathbf x - \mathbf y)
\]
There is also a conserved charge
\begin{equation}
\begin{split}
& Q  =   i  \int d^D\! x \, \left( \pi \varphi - \pi^\star \varphi^\star \right) \\
& \frac{dQ}{dt} \ = \ \{ Q, H \} = 0
\end{split}
\label{QN}
\end{equation}
and we conclude from Hamilton's equations
\begin{eqnarray}
\dot \varphi & = & \{ \varphi, H \} = \pi \\
%\label{H0-1}
%\end{equation}
%\begin{equation}
\dot \pi & = & \{ \pi, H \} = \nabla^2 \varphi - \frac{\partial U}{\partial \varphi}  
\label{H0-2}
\end{eqnarray}
that whenever the charge (\ref{QN}) has a non-vanishing value $q$, the field $\varphi$ must vary with time.  
But not all $q\not=0$ minimum
energy configurations are Q-balls, or time crystals. For (\ref{SQ1}), (\ref{QN}) to support a
 {\it time crystalline Q-ball}  the potential $U(\varphi)$ needs to be chosen in a particular fashion.

\begin{figure}[tbp]
\centering % \begin{center}/\end{center} takes some additional vertical space
%\includegraphics[width=.45\textwidth,trim=0 380 0 200,clip]{figure-1.pdf}
%\hfill
\includegraphics[width=.50\textwidth,origin=c,angle=0]{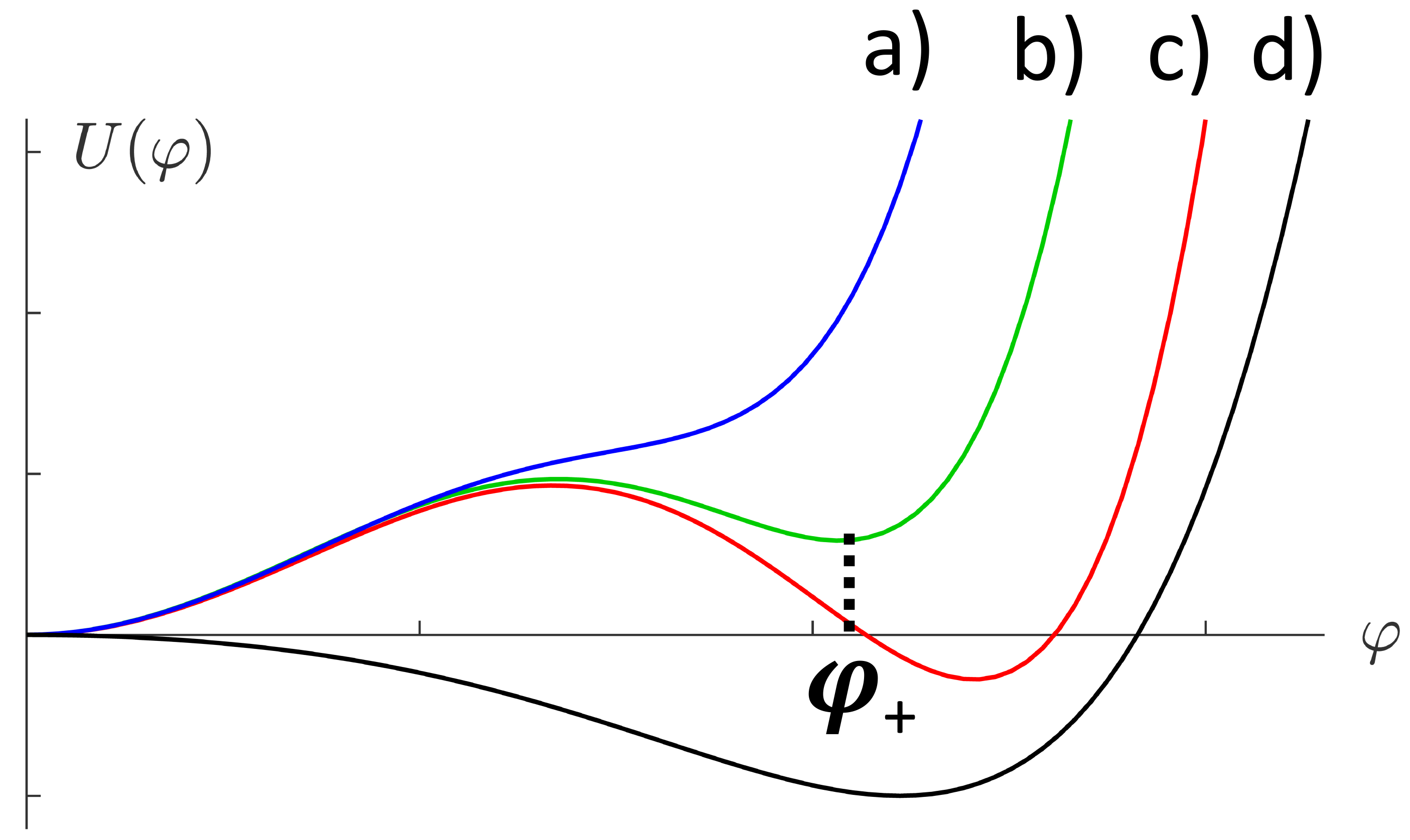}
% "\includegraphics" is very powerful; the graphicx package is already loaded
\caption{\label{fig-4} The potential a) has a single global minimum, at the origin. The potential b) has a global minimum at origin, and an additional local
minimum at non-vanishing value $\varphi_+$. 
The potential c) has a local minimum at the origin, and a global minimum at a non-vanishing value of $\varphi$. 
The potential
d) has a local maximum at the origin, and a global minimum at a non-vanishing value of $\varphi$. 
As shown in \cite{Coleman-1985} a  potential with the profile b) can support Q-balls in $D=3$}%brings about }
\end{figure}

In \cite{Coleman-1985}  a Q-ball potential with the following properties was introduced, in the case of $D=3$.
There should be no spontaneous symmetry breaking, the origin $\varphi_0 = 0$ 
should be the global minimum of   $U(\varphi)$. In addition  $U(\varphi)$ should have a local minimum at $\varphi_+ \not= 0$ and
in the limit $|\varphi|\to \infty$
the value of $U(\varphi)$ should go to infinity.  
Accordingly,  the profile of $U(\varphi)$ should resemble the potential  b) in Figure \ref{fig-4}.  
%
%
%                   FIGURE-4
%
%
%
A Q-ball can then form,  when  for small $|\mathbf x | $  the field $\varphi(\mathbf x,t)$  acquires a value close to $\varphi_+$ and  
%%this local minimum of $U(\varphi)$ needs to  have  a lower {\it total} energy value than $\varphi_0$  when $|\mathbf x | $ is small.   
approaches the global minimum $\varphi_0 = 0$ of  $U(\varphi)$ for large  $|\mathbf x | $. The Q-ball is
a stable finite energy spherical domain wall that separates the two ground states,   it is a non-topological soliton that 
interpolates between the ground state $\varphi_+$  in its interior and the ground state $\varphi_0$ in its exterior, 
with a profile and time dependence specified by the details of the action.

We are interested in a {\it time crystalline Q-ball}. This is a Q-ball that is a minimum of the Hamiltonian energy  in (\ref{SQ1})
with charge $q\not=0$. We  search for it using the general formalism of Section 2; from (\ref{SQ1}), (\ref{QN}) the Hamiltonian (\ref{crit}) is 
%In the literature one often chooses
%\begin{equation}
%H(\varphi,\pi) \ =  \  \pi \pi^\star + |\nabla \varphi |^2  + 
%m |\varphi |^2 - g |\varphi |^4 + \mu  |\varphi |^6
%\label{Uphi}
%\end{equation} 
%where the parameters $m, \ g$ and $\mu$ are all positive. 
%
%
%By rescaling the variables and coordinates we can write the action as
%\begin{equation}
%S \ = \ \int dt d^Dx \left\{ \pi \dot \varphi + \pi^\star {\dot \varphi}^\star - \pi \pi^\star - |\nabla \phi |^2  - U(|\phi|) \right\} 
%%- \frac{1}{2} m^2 |\phi |^2 + \frac{1}{4} \lambda  |\phi |^4 - \frac{1}{8} \sigma |\phi|^6 \right\}
%\label{SQ2}
%\end{equation}
\begin{equation}
%\begin{split}
H_\lambda  =  
\int d^D\! x \, \left\{ \pi \pi^\star +  |\nabla \varphi |^2  +  
|\varphi |^2 -  |\varphi |^4 +  \mu |\varphi |^6 \right\} 
+  i  \lambda \left\{ \int d^D\! x\, \left(  \pi \varphi - 
\pi^\star \varphi^\star \right) - i q \right\}
%\end{split}
\label{SQfin}
\end{equation}
where the Lagrange multiplier $\lambda$ enforces  
the conserved charge (\ref{QN}) to have the prescribed value $q\not=0$. 
When $\pi_{cr}, \varphi_{cr}, \lambda_{cr}$ solve the pertinent equation (\ref{Glambda}) so that
$\pi_{cr}, \varphi_{cr}$ minimize the energy in (\ref{SQ1})  with a Q-ball like domain wall profile, so that $\lambda_{cr}\not=0 $, we have a 
time crystalline Q-ball with time evolution given by (\ref{creq}),  
\begin{equation}
\begin{split}
& \dot\varphi \ = \ i \lambda_{cr} \varphi  \hspace{2.2cm}  {\rm with} \ \ \  \varphi(0, \mathbf x) = \varphi_{cr}(\mathbf x) \\
& \dot\pi \ = \ - i \lambda_{cr} \pi  \hspace{2.0cm}  {\rm with} \ \ \ \pi(0, \mathbf x) = \pi_{cr}(\mathbf x) 
\end{split}
\label{tcequ}
\end{equation}
%with the initial conditions $\varphi(0, \mathbf x) = \varphi_{cr} $ and $\pi(0, \mathbf x) = \pi_{cr} $.

To numerically construct explicit examples of time crystalline Q-balls, we introduce 
a discrete variant of (\ref{SQfin}) on a one dimensional lattice with $N$ sites: We discretize the gradient 
and we redefine the variables  so that we are left with  the following  version of (\ref{crit})
\begin{equation}
\begin{split}  
& H_\lambda   = H + \lambda (Q-q) \\
&  = \!\sum\limits_{k=1}^{N}\!  \left\{ \pi_k \pi^\star_k- \epsilon (\varphi_{k+1}^\star \varphi_k + \varphi_{k+1}\varphi_k^\star) +|\varphi_k| ^2 - |\varphi_k|^4 
+ \mu |\varphi_k|^6 \right\} 
+ i \lambda \{ \sum_{k=1}^N  (\pi_k \varphi_k- \pi_k ^\star \varphi_k ^\star)  + i q \}
\end{split}
\label{discH}
\end{equation}
where the second term  emerges from the cross-term of 
the discretized gradient; we set $ \varphi_{N+1} = 0$. We note that  (\ref{discH})  has the two-fold  reflection symmetry 
$k \to N+1-k$ that corresponds to a lengthwise reflection of the chain.
In a conceivable physical application, the values ($\epsilon, \mu$)  then characterize  a particular physical  environment and the values $q$ of the
conserved charge
specifies the presymplectic slice $\mathcal M_g$ in (\ref{Gg}). Accordingly, we search for time crystalline Q-balls with length $N$ and
characterized by the value of $q$, in the environment of parameter values ($\epsilon, \mu$).

The critical points ($\varphi_{k,cr}, \pi_{k,cr}, \lambda_{cr}$) of (\ref{discH}) are solutions to 
the ensuing equations (\ref{Glambda}),
\begin{equation}
\begin{split}
&  \pi_k =  i \lambda \varphi^\star_k  \\
& 
\epsilon( \varphi_{k-1} + \varphi_{k+1}) -  \varphi_k + 2 |\varphi_k |^2 \varphi_k - 3 \mu |\varphi_k|^4 \varphi_k = -  i \lambda \pi^\star_k \ \ \ \ \ (\varphi_0 =0)  \\
& \pi_k \varphi_k- \pi_k ^\star \varphi_k ^\star  = - i q 
\end{split}
\label{deom}
\end{equation}
and we  follow the steps  of Section 2 to numerically search for  a time crystalline Q-ball.  We use
constrained nonlinear optimization algorithm \cite{Byrd-2000} that we implement
using {\it MATLAB} to mimimize the energy
\begin{equation}
\min_{\pi_k,\varphi_k \in \mathbb R} \left\{  \sum\limits_{k=1}^{N} \pi_k \pi^\star_k- \epsilon (\varphi_{k+1}^\star \varphi_k + \varphi_{k+1}\varphi_k^\star) +|\varphi_k| ^2 - |\varphi_k|^4 
+ \mu |\varphi_k|^3 \right\}  
\label{minvar}
\end{equation}
subject to the condition
\begin{equation}
\sum_{k=1}^N  ( \pi_k \varphi_k- \pi_k ^\star \varphi_k ^\star ) =   - i q
\label{minQ}
\end{equation}
The minimization gives us critical values $\pi_{k,cr}$ and $\varphi_{k,cr}$. We ensure that these correspond to a global, not just a local, 
minimum of energy (\ref{minvar}) by using a large pool of randomly generated initial values in the optimization algorithm. 
We then evaluate $\lambda_{cr}$ from the first equation (\ref{deom}),
\begin{equation}
\lambda_{cr} =  i \, \frac{\pi^\star_{k,cr}}{ \varphi_{k,cr}  }  
\label{lcm}
\end{equation}
For a time crystal $\lambda_{cr}\not=0$ and in particular we verify  that $\lambda_{cr}$ has a value which is independent 
of the index $k$ which serves as a consistency check of our minimization result.

\begin{figure}[tbp]
\centering % \begin{center}/\end{center} takes some additional vertical space
%\includegraphics[width=.45\textwidth,trim=0 380 0 200,clip]{figure-1.pdf}
%\hfill
\includegraphics[width=.550\textwidth,origin=c,angle=0]{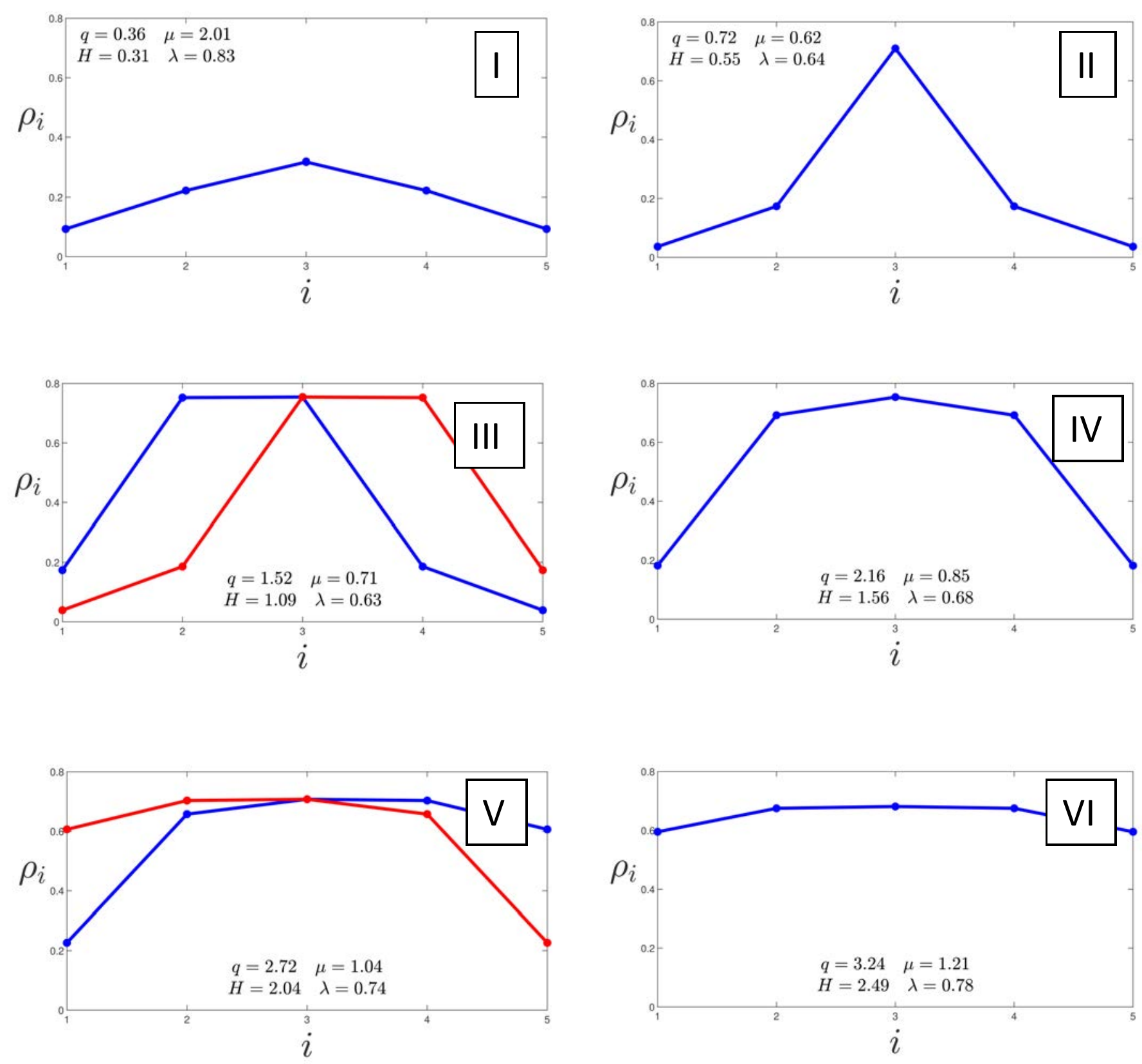}
% "\includegraphics" is very powerful; the graphicx package is already loaded
\caption{\label{fig-5} Values of order parameter (\ref{orpar1}) for minimum energy configuration (\ref{minvar}), (\ref{minQ}) with 
different representative parameter values ($q,\mu$) for a chain with 
$N=5$ sites and with $\epsilon = 1/8$.
Panels I and VI do not describe a Q-ball; there is no domain wall.  Panels II-IV are time crystalline Q-balls with two domain walls;
the blue and red profiles in Panel III are reflection symmetric under $k \to 6-k$. In
Panel V we have a pair of reflection symmetric  single domain wall time crystalline Q-balls. 
%Note that all energy minima are reflection symmetric, under $k \to 6-k$. 
}
%brings about }
\end{figure}

\begin{figure}[tbp]
\centering % \begin{center}/\end{center} takes some additional vertical space
%\includegraphics[width=.45\textwidth,trim=0 380 0 200,clip]{figure-1.pdf}
%\hfill
\includegraphics[width=.50\textwidth,origin=c,angle=0]{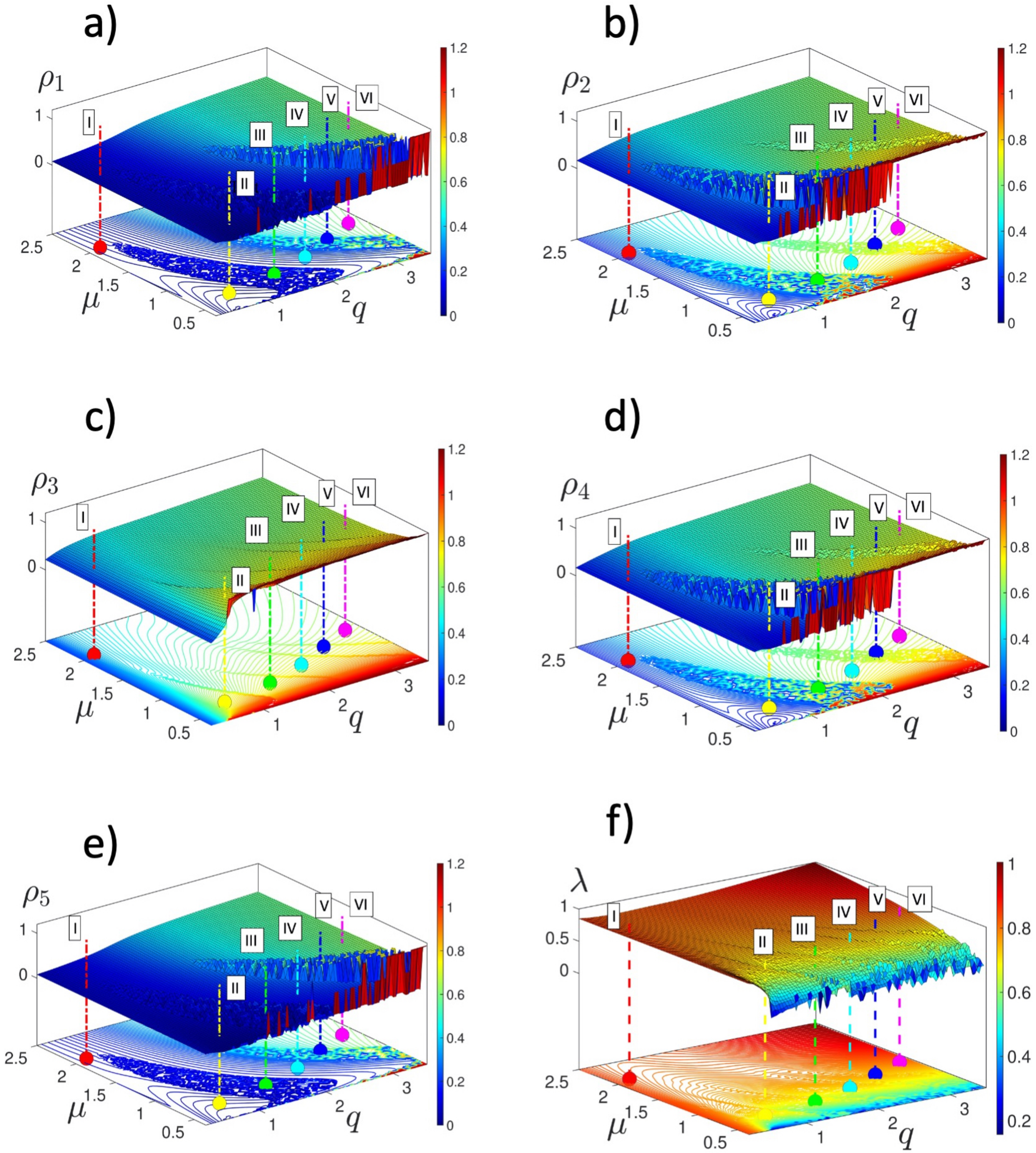}
% "\includegraphics" is very powerful; the graphicx package is already loaded
\caption{\label{fig-6} The landscape of time crystalline Q-balls in terms of ($q,\mu$). The Panels a)-e) show the landscape for the five 
order parameters $\rho_1 ... \rho_5$, respectively.  In each panel the $\rho_k$ of the six energy minima shown in Figures \ref{fig-5} are identified.
The roughness of the landscape is due to the $k \to 6+1-k$ reflection symmetry: Due to this symmetry of (\ref{minvar}),
a time crystalline Q-ball corresponds to a double degenerate ground state energy. The roughness arises since at each point
the landscape is constructed from a randomly chosen initial configuration, to display the double degeneracy.
Panel f) then shows the $|\lambda_{cr}|$ values, evaluated from (\ref{lcm}).
}
\end{figure}

When the number of variables increases, constrained nonlinear optimization becomes very quickly highly time consuming,   
thus we limit our search of  time crystalline Q-balls  to a relatively small number of lattice sites $N$. 
The Figures \ref{fig-5} and \ref{fig-6} sketch out our results with $N=5$ and  with fixed 
$\epsilon = 1/8$.  We select  $\mu \in [ 0.15,2.5]$  and the conserved charge
$q\in \ [0,4]$ as we have found that in this range of ($\mu,q$)  values time crystalline 
Q-balls can be found. For numerical simulation, we divide [$\mu,q$]  into a 100x100  lattice 
and for each pair ($\mu,q$)  we perform 100 independent
minimizations, starting from randomly chosen initial values. In the Figures \ref{fig-5} and \ref{fig-6}  we use 
\begin{equation}
\rho_k = \sqrt{ \varphi^\star_{k,cr} \varphi_{k,cr} } \ \ \ \ \ k=1,...,5
\label{orpar1}
\end{equation}  
%
%
%                   FIGURE-5
%
%
%
as the order parameter. For the corresponding momentum order parameters 
\[
\sigma_k = \sqrt{ \pi^\star_{k,cr} \pi_{k,cr} } \ \ \ \ \ k=1,...,5
\]
the results are very similar since according to  (\ref{lcm}) $\sigma_k/\rho_k = |\lambda_{cr}|$ for all $k$.  

\vskip 0.2cm

Figure  \ref{fig-5}  panels I-VI show examples of the energy minima that we find, for different parameter values. 
The six examples we depict are 
generic and chosen to describe what we find in the range of parameters we investigate.
The panels are ordered according to increasing conserved charge value $q$.
%All the solutions can be related to an effective potential profile that appears in Figure 4.
%All the energy minima reflect the two-fold $k \to N+1-k$ reflection symmetry of the Hamiltonian energy function (\ref{minvar}).

\vskip 0.3cm
$\bullet$ The Panel I characterizes  the small-$q$ and large-$\mu$ region. We propose 
that this configuration corresponds to a situation akin that shown
in Figure \ref{fig-4} profile a):  In terms of such an {\it effective} potential energy interpretation, there is only the global minimum that is 
located at  $\rho \approx 0$ ({\it i.e.} very small). In particular, there is no domain wall.

$\bullet$ The  Panel II shows how the minimum energy configuration in Panel I evolves when we increase the value of $q$, and lower the value of $\mu$:
The value of $\rho_3$ at the center of the chain increases until it reaches $\rho_3 \approx 0.7$ as shown in the Panel II. In terms of Figure \ref{fig-4} we propose the interpretation that the {\it effective} potential energy approaches a transition between the profiles a) and b). That is, 
a local minimum emerges in the {\it effective} potential energy, near $\rho\approx 0.7$: The profile  resembles a pair 
of domain walls, close to each other.

$\bullet$ In Panels III and IV we  increase the value of $q$, with only small changes in $\mu$.  In both Panels the
$\rho$  values move  back and forth between $\rho \approx 0$ and $\rho \approx 0.7$, when we move along the chain. Note that in 
Panel III the energy minimum has a two-fold degeneracy, corresponding to the $k \to 5+1-k$ reflection symmetry of  (\ref{minvar}) while in Panel IV 
the reflection symmetry becomes restored as the energy minimum is reflection symmetric.  
In terms of the {\it effective} potential energy description of Figure \ref{fig-4} we propose 
that these Panels correspond to the case b) with  only a small energy difference between the two ground states:
The profiles describe two domain walls that move away from 
each other as $q$ increases.

$\bullet$ In the  Panel V we increase the value of $q$ further while  the value of $\mu$ is more or less intact. The energy minimum 
interpolates between $\rho \approx 0$ and $\rho \approx 0.7$ and we observe two-fold
degeneracy due to reflection symmetry.  In terms of the {\it effective} potential energy 
description of Figure \ref{fig-4} we propose that this Panel corresponds to the case
between b) and c) with the qualitative feature of the Q-ball  potential energy of
 \cite{Coleman-1985,Lee-1992,Volkov-2008}: Each of the degenerate energy minima describes a single domain wall.

$\bullet$ Finally, Panel VI represents the region of large $q$ but not so small $\mu$ values.   Now $\rho$ appear to 
reside in the ground  state with $\rho \approx 0.7$.  The energy minimum resembles 
the scenario d) of Figure \ref{fig-4} and there is no domain wall.

\vskip 0.2cm
The Figure \ref{fig-6} Panels a)-e) show the entire ($q,\mu$)-landscape of $\rho_k$ minimum energy 
solutions, in the range of ($q,\mu$) values that we have studied; the Panels show the landscape of minimum energy $\rho_k$ values
both as a
surface map and as a contour map. In each of the individual Panels  we have marked the six solutions that we have detailed 
in Figures \ref{fig-5}.
%
%
%                   FIGURE-6
%
%
%

The Panel f) of the Figure \ref{fig-6} shows the values  of $\lambda_{cr}(q,\mu)$, evaluated from  (\ref{lcm}).  The Panel shows that 
each of the energy minima in Figures \ref{fig-5}  have $\lambda_{cr} \not=0$.  
Thus they are all time dependent energy minima, each determines a time dependent, time crystalline
symmetry transformation (\ref{creq}) of the Hamiltonian, of the form
\[
\begin{split}
& \varphi_k (t) = \varphi_{k,cr}(\mu,q)  \, e^{i\lambda_{cr}(\mu,q)  t} \\
& \pi_k (t) = \pi_{k,cr} (\mu,q) \, e^{-i\lambda_{cr}(\mu,q) t}
\end{split}
\]
For each lattice site $k$ this describes uniform rotation on the ($\varphi_k, \pi_k$) plane with angular 
velocity $\lambda_{cr}(\mu,q)$ independently of the $k$-value.

From Figures \ref{fig-5} and \ref{fig-6} we confirm that the energy minima shown in Panels I and VI of Figure \ref{fig-5} have no 
domain walls.  Thus they are
not Q-balls, as defined in \cite{Coleman-1985}.  Accordingly we do not consider them to be
time crystals, either:  These two Panels 
describe minimum energy configurations with (essentially) uniform $\rho_k$-values, there is 
no interpolation between different ground states as there is no domain wall.   Thus we can smoothly deform them to the ensuing 
ground state, with $q=0$.
%The situation corresponds to the profiles a) and d) in  Figure \ref{fig-4}.  
From the point of view of our general formalism in Section 2, in these two cases the corresponding orbit $\mathcal M_g$ is 
path connected to a time independent minimum energy configuration with $q=0$.
Moreover,  Figures \ref{fig-6}  show that we can path connect these two energy minima to each other, continuously in the ($q,\mu$) plane, 
without encountering any domain wall region in between, by a uniform increase/decrease of the $\rho_k$ values. 

The situation is different  in the cases shown in Panels II-V of Figure \ref{fig-5}. In each of these Panels,   
the minimum energy configuration has a domain wall profile. 
Accordingly, in line with \cite{Coleman-1985}  we interpret these four cases of energy minima as genuine time crystalline Q-balls.  
Note that the example in Panel V is a time 
crystalline Q-ball with a single domain wall,   in line with the Q-ball constructed in \cite{Coleman-1985}. The examples in Panels III and IV display  
time crystalline Q-balls with a pair of domain walls, that move  and come together as shown in Panel II, when $q$ decreases.  
We remark that both spherical and toroidal  
Q-balls have been described in the literature \cite{Volkov-2008}.

More generally, we deduce that  time crystalline Q-balls, {\it i.e.} energy minima with a domain wall structure,  
exist for those ($q,\mu$) values
where the landscapes in Figures \ref{fig-6}  exhibit roughness: The roughness is due to the presence of domain walls, it denotes
regions where we find a doubly degenerate energy  ground state. This is a consequence of the $k\to N+1-k$ reflection symmetry of the
Hamiltonian, the double degeneracy  
of the ground state is necessary for a  domain wall to be present.

\section{\large Conclusions}

We have shown, both using the general formalism of  geometric mechanics  and in terms of explicit examples, that 
classical Hamiltonian time crystals do exist and can be found in Hamiltonian systems with  conserved charges.
The {\it No-Go} arguments  only apply on a phase space with a symplectic structure, but in the case of a time
crystal the phase space is pre-symplectic.  
In particular, our general formalism establishes that a symmetry is necessary for the existence of
a time crystal. Thus the provenance of a Hamiltonian time crystal lies in the general phenomenon of spontaneous symmetry breaking that
now takes place in a dynamical context.

\acknowledgments

The research of AA was supported in part by the National Center for Competence in Research (NCCR) SwissMAP and by the grants number 178794 and 178828 of the Swiss National Science Foundation. AA is grateful for hospitality to the CRM of the University of Montreal and to the IHES where this project was developed.
The work by JD and AJN is supported by the Carl Trygger Foundation and by the Swedish Research Council (VR),
JD and AJN also acknowledge collaboration under COST Action CA17139. JD and AJN thank Frank Wilczek and Xubiao Peng for discussions.


\end{document}